# Unsupervised Machine Learning for Experimental Detection of Quantum-Many-Body Phase Transitions

Ron Ziv[1], David Wei[2,3], Antonio Rubio-Abadal[2,3], Daniel Adler[2,3,4], Anna Keselman[5], Eran Lustig[6], Ronen Talmon[1], Johannes Zeiher[2,3,4], Immanuel Bloch[2,3,4] and Mordechai Segev[1,5]

(1) Department of Electrical Engineering, Technion, Haifa 3200003, Israel
(2) Max-Planck-Institut für Quantenoptik, 85748 Garching, Germany
(3) Munich Center for Quantum Science and Technology (MCQST), 80799 Munich, Germany
(4) Fakultät für Physik, Ludwig-Maximilians-Universität München, 80799 München, Germany
(5) Department of Physics and Solid State Institute, Technion, Haifa 3200003, Israel
(6) Edward L. Ginzton Laboratory, Stanford University, Stanford, 94305, CA, USA
*msegev@technion.ac.il

**Abstract**

Quantum many-body (QMB) systems are generally computationally hard: the computing resources necessary to simulate them exactly can often exceed the existing computation resources by orders of magnitude. For this reason, Richard Feynman proposed the concept of a quantum simulator: quantum systems engineered to obey a prescribed evolution equation and repeating the experiment multiple times. Experimentally, however, as we explain below, the vast majority of observables describing the system are inaccessible. Thus, while Feynman's idea addresses the problem of simulating quantum dynamics, it leaves unsolved the equally fundamental problem of inferring the underlying physics from the limited observables accessible in experiments. Indeed, many complex phenomena associated with QMB systems remain elusive. Perhaps, the most important example is identifying phase transitions in QMB systems when no simple order-parameter exists, which poses major challenges to this day. Complicating the problem further is the fact that, in most cases, it is impossible to learn from numerical simulations, as the underlying systems are often too large to be computable, and small QMB can show strong finite size effects, masking the presence of the transition. Here, we present an unsupervised machine learning approach to study QMB experiments, specifically aimed at detecting phase transitions and crossovers directly from raw experimental measurements. We demonstrate our methodology on systems undergoing Many-Body Localization cross-over and Mott-to-Superfluid phase-transition, showing that it reveals collective phenomena from the very partial experimental data and without any model-specific prior knowledge of the system. This approach offers a general and scalable route for data-driven discovery of emergent phenomena in complex quantum many-body systems.

**Introduction**

Phase transitions are physical processes where a system containing many constituents undergoes a fundamental change in behavior due to a small change of its parameters. Examples include transitions between metal to insulator [1], gas to liquid [2], topological phase transition [3], and more [4]. Deciphering the conditions of such transitions, i.e., the specific regimes under which a transition occurs, has enabled the discovery of numerous fundamental phenomena and the engineering of matter to exploit phase-dependent properties.

Generally, phase transitions occur not only in the classical regime, driven by classical thermal fluctuations between the physical constituents of a system, but also in quantum many-body systems [5] driven by quantum fluctuations, where the task of identifying the phase transition can often be challenging. QMB systems are systems involving many quantum particles (or quantum degrees of freedom), and are described by a high-dimensional Hilbert space whose dimension scales exponentially with the number of particles. Consider the simplest case of $n$ quantum particles where each particle has just 2 possible states. In this simple case, each particle can be represented by a qubit, the dimension of the density matrix (describing the quantum information in the system's state) is $2^n$, and the number of elements in this matrix is $\sim 2^{2n}$. Obviously, even for moderate $n$, extracting all the elements of the underlying density matrix from experiments becomes impossible even for efficient quantum state tomography methods [6,7]. The problem of identifying phase transitions can be equally severe: many current QMB experiments are fundamentally limited in the type of observables that are accessible and hence in the information they provide, in particular when only bulk quantities can be measured. Examples include the conductivity of a material or magnetic properties such as the (global) magnetization or magnetic susceptibility.

This makes the detection of phase transitions from the available partial experimental measurements extremely difficult, as even knowing the indicators or order parameters of the transition can be limited[8–11], and the accessible observables often do not indicate the presence of a QMB transition[12,13]. Finally, one could think of using numerical simulations as indicators for QMB phase transition, but in 2D and 3D QMB systems simulations are unfeasible due to the computational crux of quantum many-body systems, requiring computational resources often far beyond current capabilities of classical computers. Even if one could simulate a smaller system in 2D[14–16], the presence of a phase-transition is unlikely to be observed, because one cannot generalize from very small systems (those that can still be simulated) to the larger sizes characterizing QMB phase-transitions; as stated by Phillip Anderson 5 decades ago: "More is Different"[17].

Quantum gas microscopes[18], which are now a well-established tool in studying QMB systems, provide access to configurational snapshots, projective measurements of the on-site atom density across a lattice, effectively capturing a single many-body configuration per shot[19,20]. Even though these snapshots are often limited to the parity of the site occupancy, restricting the access to most of the elements of the density matrix, analysis of the configurational snapshots opens many new opportunities for identifying phase transitions. Recent experiments have implemented protocols and pre-readout operations to extract additional observables, such as local coherence and currents[21,22], which can further enhance the amount of information extractable from single-site snapshots.

These experiments realize Feynman's vision of the quantum simulator, systems that can reproduce the forward dynamics of complex quantum evolutions. Yet they also expose a fundamental limitation: while such simulators emulate the dynamics faithfully, they yield only partial snapshots of the underlying state, making the inverse inference of physical behavior highly nontrivial. As such, *the main problem remains: identifying phase transitions from the extremely partial data available QMB experiments is still extremely challenging*.

To some extent, the problem of identifying phase transitions also exists in some classical system. With the rise of machine learning (ML), this challenge has attracted significant attention, leading to a variety of approaches. Early efforts addressed utilized supervised learning on *simulated classical systems*, such as the pioneering work of Carrasquilla and Melko[23]. These were followed by hybrid approaches that combined supervised architectures with unsupervised discovery strategies[24,25] and by fully unsupervised methods capable of identifying phases directly from data[26] even demonstrated experimentally on topological phases[27]. In attempt to extend these ideas to quantum systems[28], several approaches used supervised[29] and hybrid[30,31] learning schemes on experimental data, marking important milestones in the field.

Yet, approaches that rely on supervised training fundamentally assume prior knowledge, hence they are limited, especially for QMB systems, where a priori knowledge of phase labels or boundaries is rarely available. To overcome this, unsupervised methods were used on numerical quantum data[32,33], and more recently unsupervised analyses of experimental quantum data have begun to emerge[34,35]. However, thus far attempts to unravel collective phenomena from experimental QMB data still often depended on implicit priors or guided choices of observables, and the use of truly model-agnostic unsupervised techniques on experimental quantum-simulation data for discovering new phenomena remains exceedingly rare. *Motivated by these developments, we seek a framework that can infer collective phenomena directly from raw experimental observables, without prior feature selection or system-specific design*.

One such phase-transition (or sharp crossover) in a QMB system, whose presence is highly debated, is many-body localization (MBL)[36,37,38]. This phenomenon is related to the disorder-induced localization proposed by Anderson in 1958[39], where the transport of a single quantum particle in a disordered lattice was analyzed, and it was found that under certain conditions transport comes to a complete halt. Already in his original paper, Anderson discussed the possibility of localization in the presence of disorder and quantum many-body interactions, but the theoretical and computational tools to study QMB localization were missing. The question may be described as follows: in quantum systems with many interacting degrees of freedom, will the presence of a random disorder lead to localization or will the system eventually thermalize? In finite 1D systems, MBL was studied analytically[40], probed numerically[41,42,43] and explored in experiments[44-46]. However, in 2D and 3D systems, even of finite size, the fate of a localization phase-transition (or crossover into a MBL glass type phase) is still debated, and the hope for answering this enigma is to somehow obtain understanding from QMB experiments despite the challenges described above.

*Here, we present an unsupervised machine learning approach for the analysis of raw snapshots from quantum many-body experiments. We specifically aim towards the detection of phase transitions (or crossovers) and the more general goal of extracting fundamental features from extremely partial experimental data, akin to QMB experiments.* Our approach is based on manifold learning, specifically an adaptation of Diffusion Maps[47] for quantum data, which learns an intrinsic representation of the data. We employ repeated measurements of the accessible observables in quantum experiments, learning directly from raw measurements without precomputing any physical quantity (such as entanglement, density matrix etc.), and without relying on any model of the system involved. Consequently, our method is agnostic to the initial state of the system or any state preparation. In fact, the approach is general and - as we show below - can be utilized in very different QMB settings and systems. To establish the correctness of our method, we first test it on numerical data from 1D quantum and from 2D classical models. Subsequently, we proceed to demonstrate our methodology on 2D experimental data from systems known as computationally hard – specifically - the Mott to

Superfluid[48] transition in equilibrium, or intractable at an exact numerical level - the MBL problem as a paradigmatic example for out-of-equilibrium many-body dynamics[49]. In doing that, we specifically concentrate on novel settings such as uniform initial states in MBL, without the judiciously prepared boundary or structural features that were traditionally used[49], to identify the MBL phase transition. Finally, we note that our method is fast and requires no training, making it ideal for on-the-fly exploration during experiments. We envision applying the ideas described here to many other QMB systems, especially where experimental measurements are highly incomplete but complex (such as many-body snapshots) and no clear metric of detection is available, as is the case of random (or unknown) initial states in many-body localization.

**Main**

Our methodology relies on a widely used data-driven method for manifold learning, known as Diffusion Maps[47] (DM), aimed at uncovering the latent structure of high-dimensional data. The central assumption underlying manifold learning is that, even though the data resides in a high-dimensional space, it is governed by a small number of effective degrees of freedom, and hence it resides on a (a priori unknown) low-dimensional manifold and may be mapped onto a low-dimensional embedding space. While this assumption may appear strong in general settings, it is often valid in physical systems - where rich correlations and constraints emerge from a limited set of underlying parameters. Moreover, experimental measurements typically probe only a narrow slice of the full configuration space, further supporting the idea that the observed data resides on a low-dimensional manifold.

Recovering this space can reveal meaningful insights into the internal structure of the system. A key strength of DM lies in its grounding in dynamical systems theory. It has been shown that, in the limit of large data, the coordinate vectors of the new embedding converge to the eigenvectors of a continuous diffusion operator defined on the underlying data manifold. As a result, the resulting embedding captures the intrinsic geometry of the data, with distances in the latent space reflecting diffusion distances - akin to how information propagates locally through the dataset. This property makes the method particularly well suited for identifying (meta)stable states and transitions between them[50]. When applied to parameterized datasets, such as those spanning different physical regimes, the embedding can naturally separate distinct regions of behavior, allowing different phases of the system to emerge in the latent space. Previous work by Lustig et al.[51] demonstrated the effectiveness of this approach in a classical photonic system, where topological phase transitions were identified using the latent coordinates from the DM embedding.

Formally, the method works as follows. Consider a dataset:

$$(1)\ S = \{S_1, S_2, \ldots S_n\}$$

where each point $S_i$ corresponds to data captured under some experimental condition and belonging to one of two phases. DM assigns to each $S_i$ a specific learned coordinate, $\Phi(S_i; S) \in \Re^d$, which can be used to classify the underlying phase. Lustig et al.[51] used this technique to identify two topological phase transitions strictly from experimental data alone, without any prior knowledge on the system. However, employing DM on experimental data obtained in QMB systems, as we do here, introduces a fundamental difference, as each experimental setting $S_i$ (e.g., a particular value of interaction strength) is associated with a quantum measurement, which yields results according to the probability distribution inherent to the description of the physical phenomena. Thus, each experimental setting corresponds not to a single observation but to a set of configurational snapshots. Formally, working with data obtained from QMB experiments, we now have

$$(2)\ S = \{S_1, S_2, \ldots S_n\}, \qquad S_i = \left\{n_1^{(i)}, n_2^{(i)}, \ldots n_m^{(i)}\right\}$$

where each experimental single-shot configurational snapshot $n_j^{(i)} \in X$ lies in some measurement space $X$ (e.g., single-site occupations).

Direct averaging over multiple single-shots (for example - each showing a 2D map of the atom density), fails to capture the full underlying geometry of the data. To illustrate this via a toy example, consider a two-site system with preference for full occupation in one of the two sites, but not both. Direct averaging to obtain the mean (expectation value) snapshot will be indistinguishable from the mean of uniform two-site system, as both will show the same uniform density occupation, losing the critical information about the fluctuations characterizing the underlying geometry of the distribution and the physics encoded in it.

Thus, we modify the traditional DM framework to handle the probabilistic nature of quantum experiments, where every experimental configuration corresponds to a set of single-snapshot measurements. We do that by learning an embedding such that each set of snapshots $S_i$, representing a specific experimental configuration, is mapped into a single point in embedded space. Thus, like the original DM algorithm, we learn $\Phi(S_i; S) \in \Re^d$, a set of embedding vectors for each of the experimental settings encompassing an ensemble of snapshots. The details of the new algorithm and its implications are given in the Methods and Supplementary sections.

First, we carry out a validation of our modified DM methodology by applying it to numerical data from systems known to undergo a phase transition. We test our methodology on a small 1D QMB system that can still be computed efficiently: the transverse field Ising model.

We simulate transverse field Ising model, which is governed by the Hamiltonian:

$$(3) \quad \hat{H} = -\sum_{j=0}^{L-1} \lambda \sigma_j^x \sigma_{j+1}^x + \sigma_j^z$$

where $\sigma_j^\alpha$ are the Pauli matrices on site $j$ and $\lambda$ is the inverse magnetic field strength. We use the Density Matrix Renormalization Group (DMRG)[42] method to compute the ground state of a chain of length $L = 32$ for various values of $\lambda$. By sampling from the computed ground state in the Z spin basis, we obtain different snapshots for each $\lambda$ value. When $\lambda \to 0$ the ground state approaches a product state, where each spin points in the positive Z direction. As $\lambda \to \infty$ the ground state is a product state pointing in the same direction along the X direction.

Testing our methodology on the simulated configurational snapshots, we learn a new coordinate representation of the data. For visualization, we plot the first embedded coordinate, $\phi_1$, against the inverse field strength $\lambda$ in Fig. 2. We observe a sharp trend change at the critical field value with two plateaus at high and low field strengths. This reveals how the latent coordinate effectively captures the different physical phases, acting as an "emergent order parameter". At this point, motivated by our numerical results on the 2D Ising model and its connection to the Curie–Weiss law in the mean-field case, we fit a hyperbolic tangent to the first embedded coordinate to extract a transition point, highlighted in the plot, to characterize the phase change. The fit points to the critical field value of $\lambda_c = 0.938(18)$, in close agreement with the known value $\lambda_c = 1$[52]. The discrepancy in the critical value may originate from finite-size effects or from the choice of hyperbolic tangent fitting. Alternative approaches, such as identifying density shifts in the embedded space or applying clustering algorithms (e.g., k-means, DBSCAN) in the embedded space, can also be used and yield similar results. Importantly, while different metrics may somewhat shift the precise transition point, they all point to the same main insight: the existence of a phase transition.

To reaffirm the ability of our methodology to identify phase transitions, we test it on another known example of a system that can be computed: the 2D classical Ising system, known to undergo a phase transition at a critical temperature value. As shown in the Supplementary Information, our methodology properly identifies the phase transition and the critical temperature value.

Gaining confidence from these results and the performance of our methodology, we proceed to the challenging case of experimental data from 2D QMB systems undergoing transitions in different settings – first equilibrium transitions, such as the Mott-to-superfluid phase transition, and subsequently dynamical or out-of-equilibrium crossover behavior, such as many-body localization scenarios (MBL). We emphasize that, while the Mott–to-Superfluid transition is challenging to simulate but is still tractable for reasonable system sizes, the MBL problem is computationally intractable for the system sizes explored here.

All the experiments described here are conducted in a quantum gas microscope. The experimental system consists of a 2D lattice of cold $^{87}$Rb bosonic atoms confined to a single plane in a vertical optical standing wave. The trapping beams introduce a harmonic confinement, which leads to a circular system size with a radius of approximately nine lattice sites. The dynamics is governed by the Bose-Hubbard Hamiltonian including a site-dependent potential:

$$(4) \quad \hat{H} = -J \sum_{\langle i,j \rangle} \hat{a}_i^\dagger \hat{a}_j + \frac{U}{2} \sum_i \hat{n}_i(\hat{n}_i - 1) + \sum_i V_i \hat{n}_i$$

Where $\hat{a}_j^\dagger$ ($\hat{a}_j$) is the bosonic creation (annihilation) operator, $\hat{n}_i$ is the local density operator of site $i$, $V_i$ is the onsite potential, and $\langle \cdot,\cdot \rangle$ denotes a summation over neighboring sites. Here, $J$ represents the hopping strength between nearest neighbor sites, and $U$ denotes the onsite interaction strength. Additional experimental details specific to each experiment are detailed in the Methods and Supplementary Information sections.

In the Mott-to-Superfluid experiment, Fig. 3, the system is initialized in a superfluid state, which is adiabatically tuned across the phase transition by varying the ratio $U/J$, followed by a measurement. In these settings, the site potential $V_i$ includes the harmonic confinement and specifically tailored, local site-blocking potentials used to realize different lattice geometries. Three distinct lattice configurations are studied: square, triangular, and Lieb lattices. As $U/J$ decreases, quantum fluctuations emerge in the form of doublon-hole pairs, captured via site-resolved parity measurements. The nearest-neighbor parity correlations grow as $\sim \left(\frac{J}{U}\right)^2$ in the Mott regime and decrease near the critical point, consistent with pair deconfinement. To probe nonlocal order, the experiment measures the Brane parity order parameter[53,54] over finite regions, a known theoretical proxy of the transition, defined as:

$$(5)\ \widehat{O}_{Brane} = e^{i\pi \delta \widehat{N}} = e^{i\pi \sum_{i\in L\times L} \delta \hat{n}_i}$$

where $\delta \hat{n}_i$ denotes the site's occupation deviation from the mean, and the sum is taken over a region of $L \times L$ sites at the center of the cloud. The Brane parity remains finite in the Mott phase and vanishes in the superfluid regime, enabling identification of the critical point. The transition behavior across geometries is compared to the latent embedding from the algorithm in Fig. 4, showing a very similar behaviour as the Brane parity observable. We obtain the critical values for the transition as $\frac{U^{ML}}{J_c} = 14.7(7), \frac{U^{ML}}{J_c} = 26(2), \frac{U^{ML}}{J_c} = 10.2(5)$ for the square, triangular and Lieb configurations respectively using the ML method. These values are consistent with the critical values extracted from the brane parity analysis[48], and are very similar to the values found through Quantum Monte Carlo (QMC) simulations for the square[55], $\frac{U^{QMC}}{J_c} = 16.739(8)$, and triangular[56] lattices, $\frac{U^{QMC}}{J_c} = 26.6(5)$. To our knowledge, no precise QMC value for the Lieb lattice has been reported to date. Comparing the values found by our methodology to those known from QMC (when the latter are known, e.g., the square lattice), indicates a difference. We believe this is a true experimental difference, (rather than one that can be attributed to the method). This discrepancy may arise from residual heating, imperfect adiabaticity, or inhomogeneities introduced by the harmonic trap, factors not present in idealized QMC simulations, which are typically performed on homogeneous systems at very low but finite temperature.

We proceed to the MBL experiments, showing results on 1D system in the supplementary, and focusing here on the challenging case of 2D systems. Fig 4 and Fig 5 present the 2D results, where the former includes data similar to the one taken using a specific initial state (a sharp edge)[49], and the latter contain new experimental results based on an unstructured initial state. Generally, MBL experiments initialized with a homogeneous distribution of atoms are considered highly challenging for identifying phase transitions / crossovers. However, they are preferrable when strong density dependences in the physical behavior of the system persist. In the experiments, disorder is introduced by projecting a site-resolved random potential onto the lattice taking $V_i = W\delta_i$, where $\delta_i$ is random disorder potential of site $i$, and $W$ is its amplitude. The system is initially prepared in a unity-filling Mott insulator state. A quantum quench is then applied by lowering the lattice depth, enabling tunneling and interactions between atoms. After the quench, the atoms are allowed to evolve freely, followed by site-resolved parity measurements. For both cases in Fig. 4 and Fig. 5 the system parameters are $\frac{U}{J} = 24(1)$. For the data presented in Fig. 4, the system is initialized with an artificial edge by removing half of the atoms from one side of the trap, creating an imbalanced state. The imbalance as a function of disorder is measured by:

$$(6) \quad Imbalance(W) = \langle \frac{N_L^W - N_R^W}{N_{total}^W} \rangle$$

where, $N_L^W(N_R^W)$ are the total atoms to the left (right) of the edge, and $N_{total}^W$ are the total atoms, per disorder strength $W$, averaged over all snapshots. The imbalance serves as a heuristic indicator of localization, reflecting the "survival" of the initial state after long evolution times. As shown in Fig. 4(B), the imbalance increases with disorder strength, consistent with the onset of localization. The insets in Fig. 4(B), depicting the mean single-shots of the spatial distribution of atoms, reveal that the initial edge shaped state is washed out when the disorder strength is increased. For exactly this crossover indication, the initial edge state was chosen and prepared at the start of experiment.

Using our methodology, we learn a new coordinate representation of the data, as shown in Fig. 4(A). For visualization, we plot the first three coordinates in this embedded space, where each point represents an ensemble of snapshots. A clear monotonic trend emerges in the embedded curve, reflecting the ordering with disorder strength. The curve reveals changes in trend and relative gaps between clusters, which correlate with disorder strength and imbalance.

Figure 4(B) shows the first embedded coordinate along with a fit to a hyperbolic tangent to extract a transition point, highlighted in the plot, to characterize the phase change. The fit points to the critical disorder strength as $W_C = 13.4(7)[J]$. This value is higher than the one reported in Choi et al.[49], but several differences can account for the discrepancy. First, as previously noted, the choice of detection criterion can shift the extracted critical value.[57] The original work identified the transition at the onset of imbalance growth; applying the same criterion to our data yields a consistent critical value. Second, our dataset spans a wider range of disorder strengths, which may smooth the latent embeddings and result in a more gradual transition, also visible in the imbalance dynamics at stronger disorder strengths.

Gaining confidence in the predictive power of the method, we now turn to the same disordered Bose-Hubbard model (Fig. 5) but importantly in a different regime of unit filling ($n = 1$) and unstructured initial state. As a result, imbalance and other heuristic measures of localization (or memory of the initial state) are no longer applicable for detecting the transition. As the images depicting the atomic density presented in Fig. 5(A) and 5(B) show, there is no indication for localization.

We apply our method to this case, following the same procedure as in Fig. 4. The results are shown in Fig. 5. The embedding (Fig. 5(A)) and the first latent coordinate (Fig. 5(B)) exhibit features similar to those of Fig. 4, indicating a clear separation of phases – despite the fact that the atom density (inserts in Fig. 5) does not show any obvious sign of localization. The data shows a slight change in the mean density across the phases. This may result from the creation of doublon–hole pairs, which, when measured via site parity, appears as a reduction in the average density. Again, a hyperbolic tangent provides a good fit for capturing the cross-over (Fig. 5(B)). Our method identifies the critical disorder strength as $W_C = 9.5(5)[J]$. Altogether, Fig. 5 exemplifies that our methodology is able to identify a strong difference in the underlying behavior of the system, hinting to either a crossover or transition in the system. We emphasize that our analysis cannot distinguish whether this change in behavior emerges due to a genuinely many-body localized (MBL) phase or to an MBL-glass phase with thermalizing instabilities occurring at long evolution times.

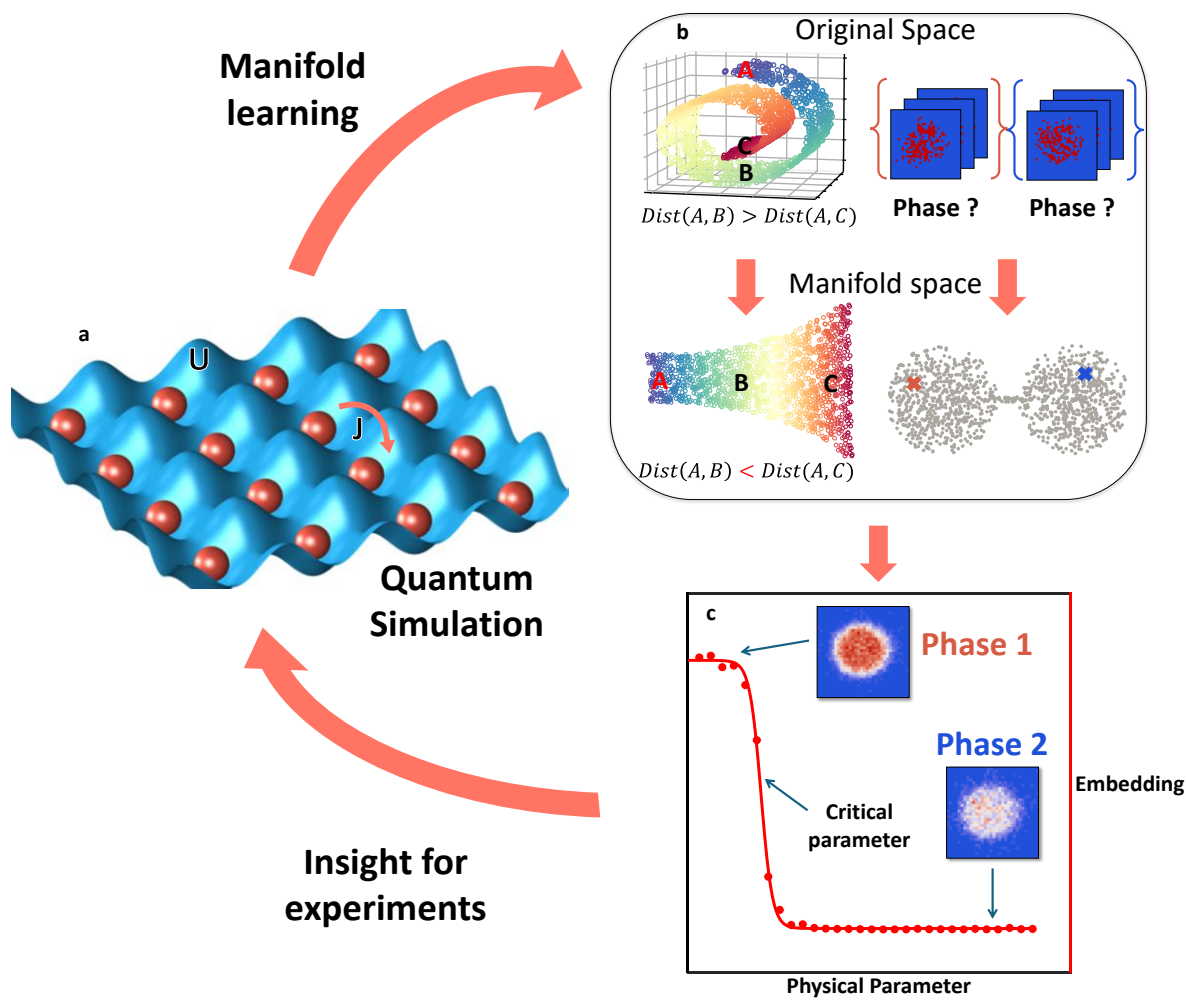

**Fig 1. Manifold learning approach for identifying critical parameters in quantum simulation experiments.**
(a) Quantum simulation platforms provide access to experimental datasets, such as configurational snapshots, probing novel quantum many-body systems. These datasets, however, are inherently limited in the information they capture. We therefore propose (b) a manifold learning approach that operates directly on raw experimental quantum data in an unsupervised manner, without prior knowledge or engineered observables. The method uncovers a data-driven low-dimensional representation that respects the intrinsic geometry of the data and enables the definition of meaningful distances between different experimental settings. In gray: an illustrative embedding of snapshot sets into a low-dimensional manifold, exhibiting a bottleneck structure and a separation between two distinct regions (left and right spheres) in latent space, signaling a qualitative change in the underlying system. This allows us to (c) identify latent structure, capture transient or crossover behavior, and detect critical parameters indicative of phase transitions. Insets show averaged snapshots from many-body localization experiments, illustrating the extracted transition behavior. The inferred critical structure can guide further exploration and experimental design.

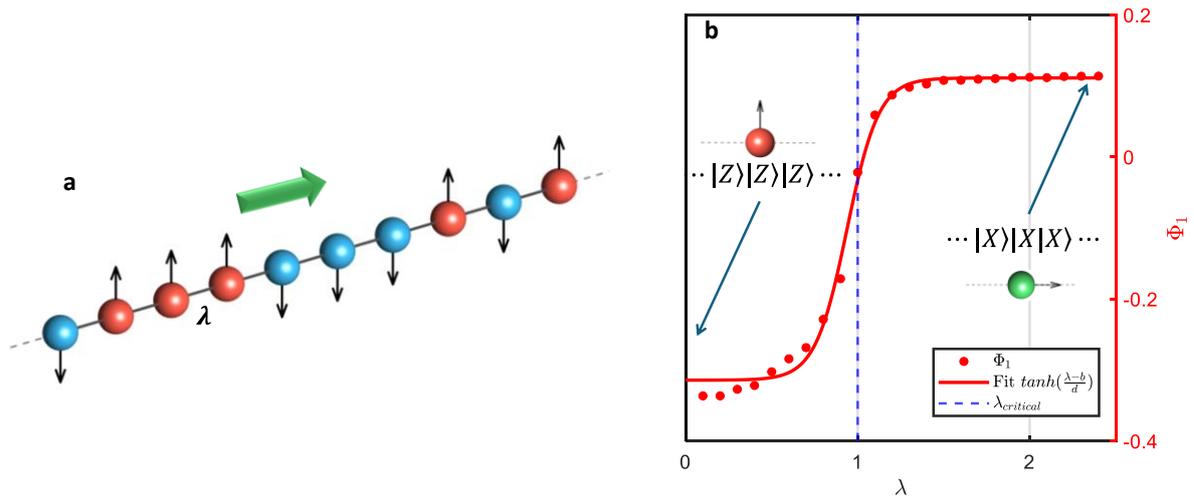

**Fig 2. Detecting phase transition from simulations of the 1D Transverse Field Ising model**. **(a)** Schematics of a spin chain under a transverse field (green arrow), depicting the transverse field Ising system. **(b)** Detecting the transition using Diffusion Maps. We generate ground states using DMRG across a range of inverse transverse field strengths $\lambda$, and sample snapshots of spin configurations from each ground state. The inserts illustrate the limiting ground state far from the quantum phase transition, highlighting the change in the ground state structure of the transverse-field Ising chain. From these snapshots we calculate the first latent coordinate, $\Phi_1$ (red circles), and identify the transition. A hyperbolic tangent fit yields a critical point, in close agreement with the known theoretical value (blue dashed line), indicating that the embedding correctly captures the underlying quantum phase structure.

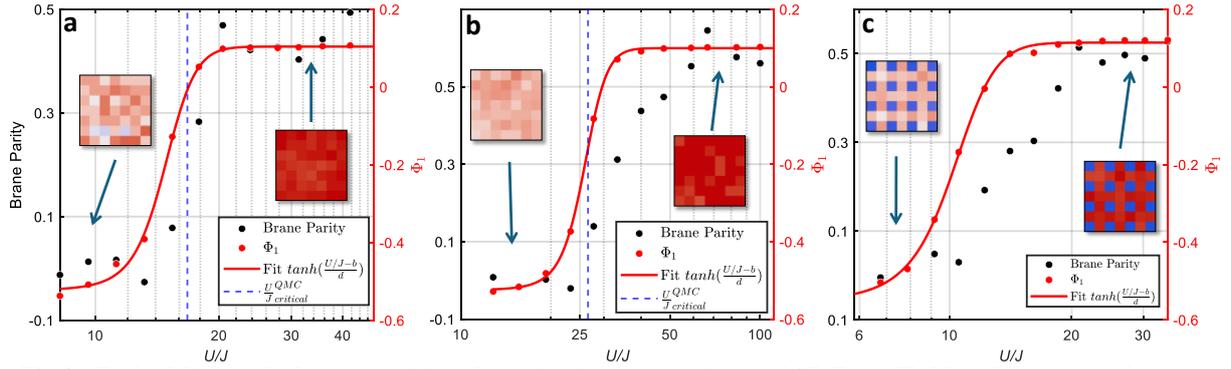

**Fig 3. Embedding and phase transition detection in an experimental 2D Bose-Hubbard system undergoing Mott to Superfluid transition under various configurations.** The leading embedding is plotted versus the Brane Parity for **(a)** square, **(b)** triangular and **(c)** Lieb lattice geometries with mean snapshot inserts. In the Mott phase, fluctuations are local, leading to a conserved particle number in a local region and finite Brane parity (Right inserts in a,b and c). In the superfluid phase, fluctuations are global and lead through the formation of doublon-hole pairs to reduction in particle number and zero Brane parity (Left inserts in a,b,c). The hyperbolic tangent fit is a heuristic fit function allowing to identify the transition for different geometries, the blue dashed line indicates the theoretical critical value from Quantum Monte Carlo numerical simulations for the square and triangular lattices. X axis is plotted in log scale.

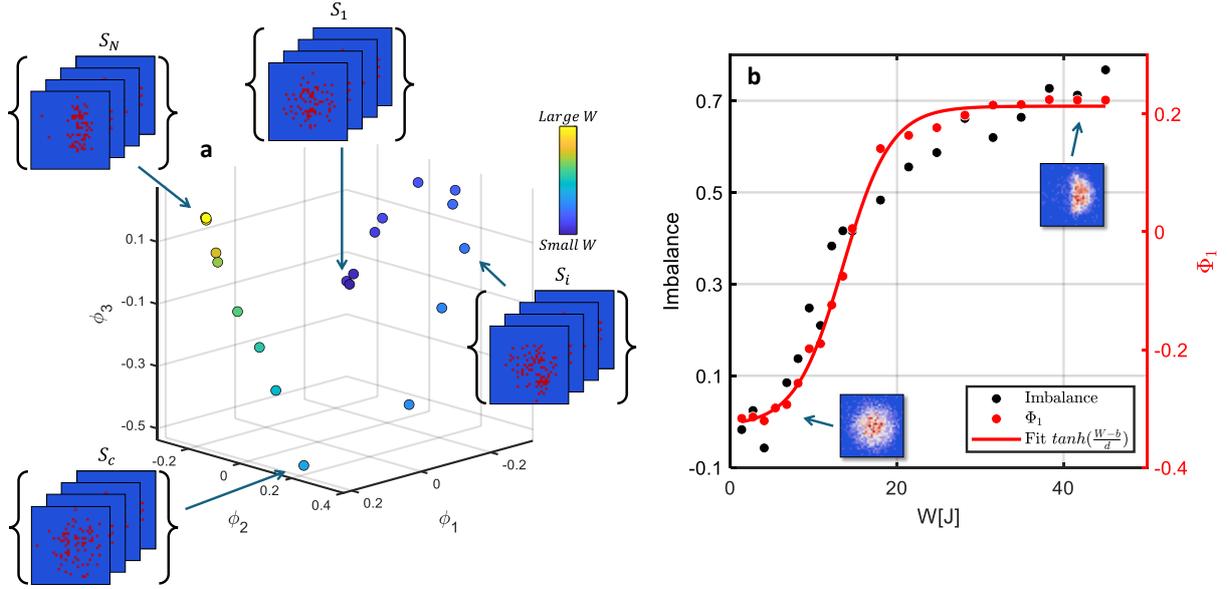

**Fig 4. Embedding and detection of the crossover in an experimental 2D disordered Bose-Hubbard system. The experiment is initiated with an edge in space (a domain wall in the atom density), which serves as an indicator for Many-Body Localization (MBL).** The inserts in A and B depict the atom density distribution (snapshots in a, mean in b) in space, as it reaches its final state. Localization occurs at sufficiently high disorder levels and is manifested by the survival of the initial edge (inserts $S_N$ in a and upper insert in b). At weak disorder levels, the edge is washed out (inserts $S_1$, $S_i$ in a and lower insert in b). Notably, the reappearance of the edge is a soft process, and gradual with disorder level (insert $S_c$ near the vicinity of the transition in a and imbalance plot in b). **(a)** Learned 3D embedding of experimental data, where each point represents an entire ensemble of raw snapshots corresponding to a specific experimental setting, color coding represents disorder strength. This unsupervised embedding captures global structure across the dataset. **(b)** The leading embedding coordinate ($\Phi_1$) correlates strongly with the initial-state imbalance, a known physical observable, suggesting that the embedding uncovers an emergent order parameter. The transition between regimes is well-captured by a hyperbolic tangent fit, reflecting the crossover behavior between distinct phases.

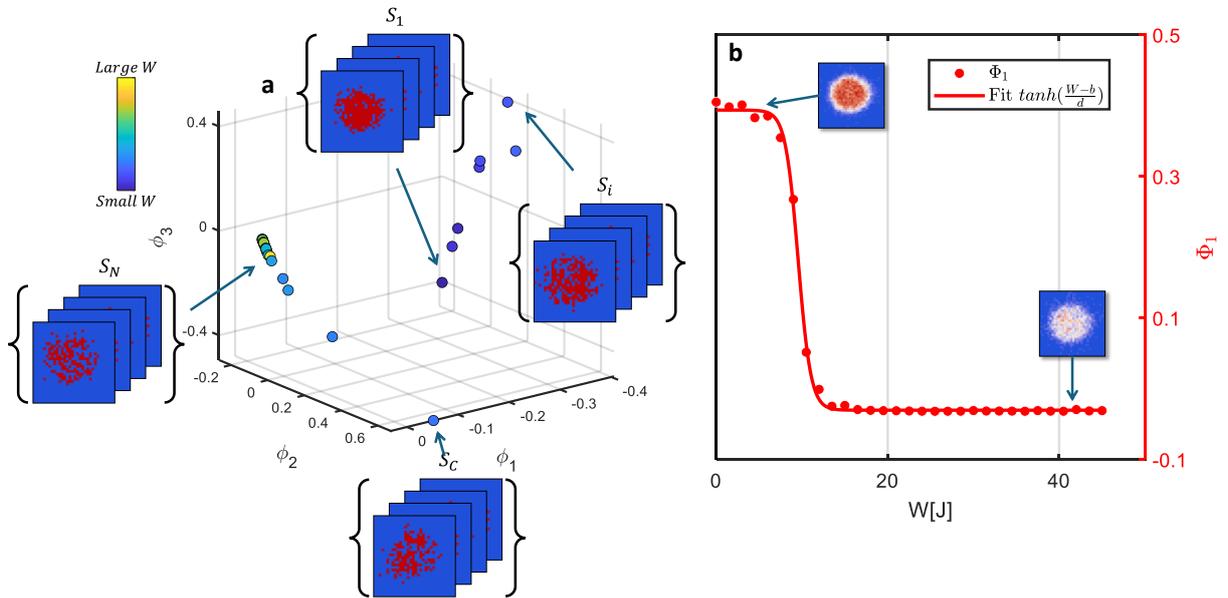

**Fig 5. Embedding and detection of the crossover in an experimental 2D disordered Bose-Hubbard system with uniform initial states**. Distinctly from the experiments of Fig. 4, here the experiment is initiated with a homogeneous atom distribution in space. As the image inserts in both a and b show, there is no clear indication of localization. **(a)** 3D embedding plot of the experimental dataset for different disorder strength. **(b)** First coordinate plot versus disorder strength, with a fit for a hyperbolic tangent, exemplifying the presence of the MBL phase-transition – despite the homogeneous initial state and not relying on any engineered state.

In summary, we have presented an unsupervised, measurement-agnostic, and state-agnostic framework for detecting quantum phase transitions or crossovers directly from raw, partial experimental data. Our approach builds on the DM algorithm, a classical tool for manifold learning, which we modify to the probabilistic setting of quantum measurements by introducing a distribution-aware kernel based on a Mahalanobis-like distance. We demonstrated for several example cases that this method can identify known phase transitions and remains effective even in challenging scenarios, such as detecting state changes in many-body localization (MBL) scenarios using uniform initial states, where traditional approaches can fail[58,59]. Crucially, the method is computationally efficient, fast, requires no GPU acceleration and typically executes in under a minute, making it well suited for rapid, exploratory analysis in QMB experiments where a clear order parameter or transition indicator may be elusive. Moreover, because the method operates directly on local measurement snapshots and makes minimal assumptions about system dynamics or observables, it is broadly applicable across quantum platforms.

Our findings show that data-driven geometric analysis can reveal collective behavior in quantum systems, even under severe measurement constraints. Without requiring access to the Hamiltonian, specific observables, preparation of a predesigned initial state, or any prior information, our method extracts useful (and potentially hidden) structures directly from raw experimental data. These findings are empirically validated on simulated and experimental datasets across diverse quantum systems, demonstrating robustness; however, the precise connection to quantum phase transitions, beyond detection of strong hidden structural changes in the data, remains an important open question.

Looking ahead, we envision integrating this framework into adaptive experimental protocols, enabling real-time detection of phase transitions and efficient mapping of complex quantum many-body dynamics. Naturally, many theoretical questions remain open, perhaps the most important one being the ability to extract and validate physical features of the different phases in models whose outcome is yet unknown and guide potential hypothesis, as exemplified by the uniform MBL case. Still, our approach provides a practical path toward scalable, model-independent analysis of quantum many-body systems, and of the discovery of new phenomena guided by machine learning.

## Methods

**Diffusion Maps for Quantum Data**

As explained in the main text, we modified the traditional DM algorithm to capture the statistical nature of data from QMB experiments. This is achieved by defining a Mahalanobis-like distance between distributions, allowing for structure-aware comparison between parameter settings. Specifically, we define the distance:

$$(3) \quad d^2(z_i, z_j) = \frac{1}{2}(z_i - z_j)^T (C_i^{-1} + C_j^{-1})(z_i - z_j)$$

where $z_i = \frac{1}{m}\sum_{j=1}^{m} p_j^{(i)}$ is the empirical mean and $C_i = \frac{1}{m-1}\sum_{j=1}^{m}\left(p_j^{(i)} - z_i\right)\left(p_j^{(i)} - z_i\right)^T$ is the empirical covariance of $S_i$. To address noise and rank deficiency, covariance inversion is approximated using truncated singular value decomposition; see the Supplementary Material for details. A similarity function $f(\cdot)$ then gives the kernel:

$$(4) \quad K(z_i, z_j) = f\left(d^2(z_i, z_j)\right)$$

This kernel forms the basis for constructing the DM embedding. A Gaussian kernel is commonly used, with its width serving as a free parameter that defines the similarity length scale between points. As a key modification, we apply a weighted wavelet transform to each snapshot prior to computing the Mahalanobis distance. This preprocessing step, related to the earth mover's distance[60], was previously shown to enhance the DM performance[61], and we find that it is critical for successful detection of the MBL problem (see Supplementary Materials for details).

Unlike the Euclidean distance, which treats all directions in the measurement space equally, the Mahalanobis distance accounts for local variance. By incorporating the covariances $C_i$ and $C_j$, the distance reflects the distributional geometry of the ensemble of each setting. This leads to a more faithful representation of the underlying parameter space and makes the approach particularly well suited for quantum settings where the experimentally accessible quantities are often very partial and the measurements have statistical distribution. As an alternative, one may embed each individual snapshot and aggregate in the latent space, an idea we have suggested in 2022[62]. However, such approach is computationally highly intensive, especially when dealing with hundreds of snapshots per setting, making it not compatible with large QMB experiments. In contrast, the Mahalanobis-based method, which we employ here, offers a substantial computational advantage, typically running in under one minute per dataset, making it viable for on-the-fly experimental analysis. In the Supplementary Material, we describe the full details of the modified methodology.

**Experimental system**

The experimental data in the main text was taken on a quantum gas microscope using bosonic rubidium-87 atoms to realize the 2D Bose-Hubbard model (BHM). The experimental sequences started with a superfluid of about 200 atoms trapped in a harmonically confining antinode of an out-of-plane optical lattice. An additional in-plane 2D lattice with a wavelength of 1064 nm allowed us to then control the tunnel coupling $J$ of the BHM. Furthermore, a digital micromirror device (DMD) allowed us to project arbitrary site-resolved potentials onto the atoms. These could be used for both potential shaping as well as initial-state preparation.

The data related to the Mott-insulator-to-superfluid quantum phase transition has been previously published in Ref. [48]. Here we used a combination of a bow-tie lattice with 752 nm spacing and a retroreflected 1D lattice to realize the square and triangular lattice geometries. The Lieb lattice was realized by adding repulsive site-blocking potentials with the DMD at a wavelength of 670 nm. Starting with a superfluid, we adiabatically ramped up the lattice in 200 ms to a depth corresponding to the targeted BHM parameters, followed by a fast ramp up to freeze the quantum fluctuations. The site-resolved occupation parity was then read out by fluorescence imaging.

The data related to the domain-wall many-body localization dynamics was measured with the same experimental parameters as the results previously published in Ref. [49]. Here we used two orthogonal retroreflected 1D lattices to realize a square lattice with 532 nm spacing. Starting with the superfluid, we ramped up the lattice adiabatically to prepare a unit-filled Mott insulator. Using the DMD with light at 787 nm and our microwave addressing technology, we then initialized the system in a domain-wall configuration. Then the DMD pattern was changed to project a shot-to-shot random disorder potential onto the lattice sites. To initiate the dynamics, the lattice depth was lowered in 5ms to the value corresponding to the targeted BHM parameters. For occupation parity detection, the lattice was quickly ramped up and the atoms read out by fluorescence imaging.

The unit-filling many-body localization dynamics data is not based on any previous publication. The sequence is identical to the domain-wall case, while keeping the disorder pattern the same throughout the different shots and skipping the domain-wall preparation step. As a consequence of this, the local density is twice as high as in the domain-wall experiments.

## Acknowledgements

We acknowledge funding by the Max Planck Society (MPG), the Deutsche Forschungsgemeinschaft (DFG, German Research Foundation) under Germany's Excellence Strategy–EXC-2111–390814868 and through the DFG Research Unit FOR 5522 (project-id 499180199). We also received funding under the Horizon Europe programme HORIZON-CL4-2022-QUANTUM-02-SGA via the project 101113690 (PASQuanS2.1).